\documentclass[aps,prl,twocolumn,nofootinbib,superscriptaddress]{revtex4-1}
\usepackage[american]{babel}
\usepackage{graphicx}  
\usepackage{dcolumn}   
\usepackage{bm} 
\usepackage{amssymb}   
\usepackage{amsmath}
\usepackage{color} 

\def\op#1{{\Hat{\mathrm{#1}}}}

\def\bra#1{\ensuremath{\langle{#1}\vert}}
\def\ket#1{\ensuremath{\vert{#1}\rangle}}
\def\bracket#1#2{\ensuremath{%
    \langle{#1}\mkern1.2mu\vert\mkern1.2mu{#2}\rangle}}

\def\abs#1{\mathinner{\lvert#1\rvert}}

\newcommand{\be}{\begin{equation}}
\newcommand{\ee}{\end{equation}}
\newcommand{\benn}{\begin{equation*}}
\newcommand{\eenn}{\end{equation*}}

\newcommand{\beq}{\begin{eqnarray}}
\newcommand{\eeq}{\end{eqnarray}}

\def\H1{\widehat{H}_1}

\newcommand{\pd}{\partial}
\newcommand{\diag}{\mathop{\mathrm{diag}}}

\newcommand{\lb}{\left[}
\newcommand{\rb}{\right]}
\newcommand{\lp}{\left(}
\newcommand{\rp}{\right)}

\DeclareMathOperator{\sgn}{sgn}

\def\H1{\widehat{H}_1}

\begin{document}

\title{Geodesic Paths for Quantum Many-Body Systems}

\author{Michael Tomka}
\email[]{tomkam@bu.edu}
\affiliation{Department of Physics, Boston University, 590 Commonwealth Ave., Boston, MA 02215, USA}
\author{Tiago Souza}
\affiliation{Department of Physics, Boston University, 590 Commonwealth Ave., Boston, MA 02215, USA}
\author{Steven Rosenberg}
\affiliation{Department of Mathematics and Statistics, Boston University, 111 Cummington Mall, Boston, MA 02215, USA}
\author{Anatoli Polkovnikov}
\affiliation{Department of Physics, Boston University, 590 Commonwealth Ave., Boston, MA 02215, USA}

\date{\today}

\begin{abstract}
We propose a method to obtain optimal protocols for adiabatic
ground-state preparation near the adiabatic limit, extending earlier
ideas from [D. A. Sivak and G. E. Crooks, Phys. Rev. Lett. {\bf 108},
  190602 (2012)]
to quantum non-dissipative systems.
The space of controllable parameters of isolated quantum many-body
systems is endowed with a Riemannian quantum metric structure, which
can be exploited when such systems are driven adiabatically. 
Here, we use this metric structure to construct optimal protocols in
order to accomplish the task of adiabatic ground-state preparation in
a fixed amount of time.
Such optimal protocols are shown to be geodesics on the parameter
manifold, maximizing the local fidelity.
Physically, such protocols minimize the average energy fluctuations
along the path.
Our findings are illustrated on the Landau-Zener model and the
anisotropic XY spin chain.
In both cases we show that geodesic protocols drastically improve
the final fidelity.
Moreover, this happens even if one crosses a critical point, where the
adiabatic perturbation theory fails.
\end{abstract}


\maketitle



\textit{Introduction.}$-$An accurate preparation of quantum states is
a fundamental requirement for the realization of emergent quantum
technologies such as quantum computers~\cite{NielsenChuang2010},
quantum sensors~\cite{qsensing}, quantum cryptography~\cite{Ekert1991}
and quantum simulators~\cite{qsim, Baumann2010, Blatt2012, Islam2013}.
To reduce the effects of noise and circumvent decoherence in such
quantum devices, it is essential to find the optimal protocol
that transforms an experimentally readily available
initial state into a desired state with high fidelity, on which the
necessary quantum manipulations are then conducted.
Quantum optimal control~\cite{Walmsley} provides powerful methods to
cope with this issue and they have been implemented in cold
atomic systems~\cite{Chu2002}, atom chips~\cite{Lovecchio2016},
superconducting quantum circuits~\cite{Goan2014} and are a vital
aspect in adiabatic quantum computation~\cite{Fahri2001}.
Optimal control algorithms for particular quantum many-body
systems have recently been developed in~\cite{Doria2011, Chamon2011},
but so far these findings are model specific.

Recently, a new general idea connecting an optimization problem and
geometry in dissipative systems was proposed in Refs.~\cite{Sivak2012,
  Zulkowski2012}. In particular, it was shown that the optimum
protocol minimizing heat along a thermodynamic path corresponds to the
geodesic associated with the metric given by the friction
tensor. These results, however, do not immediately extend to low
temperature systems, where the friction tensor vanishes and leading
non-adiabatic corrections come from virtual excitations determining
the mass renormalization~\cite{DAlessio2014}. 
  A time-optimal approach to adiabatic quantum
  computation was formulated in a differential-geometric
  framework by A.~T.~Rezakhani {\it et al.}~\cite{Rezakhani2009}.
  They demonstrated that the optimal strategy, keeping the
  evolution adiabatic, is given by a geodesic. 
  Although, in their set-up the adiabatic condition is used as a
  heuristic to define a metric tensor, and therefore there might
  exist a better definition.
%
%
%
%
%
%
%
%
In this work, we extend the ideas of Refs.~\cite{Sivak2012, Zulkowski2012, Rezakhani2009} by using the Fubini-Study quantum
metric associated with the quantum fidelity~\cite{Provost1980}. This metric equips the space of control parameters with a Riemannian structure~\cite{Zanardi2007, Kolodrubetz2013, Kolodrubetz2016}.

Let us consider a closed quantum many-body system, described by a Hamiltonian $H(\vec{\lambda}(t))$ depending on time through the control parameters,
$\vec{\lambda}(t)=(\lambda^{1}(t) , \ldots, \lambda^{p}(t))^{T}$,
where $p$ is the dimension of the parameter manifold
$\mathcal{M}$.
The problem of optimal adiabatic state preparation is then stated
as follows:
find the optimal protocol
$\vec{\lambda}_{\mathrm{opt}}(t)$
that transforms $\ket{\psi_{0}(0)}$, initial ground-state of
$H(\vec{\lambda}(0))$, to the desired state
$\ket{\psi_{0}(t_{f})}$, ground-state of $H(\vec{\lambda}(t_{f}))$.
As a measure of similarity between the evolved state
$\ket{\psi(t_{f})}$ and the target state $\ket{\psi_{0}(t_{f})}$,
we use the fidelity
\be
\mathcal{F}\big[\vec{\lambda}(t_{f})\big]
=
\abs{\bracket{\psi(t_{f})}{\psi_{0}(t_{f})}}^{2}.
\ee
The task is therefore to find $\vec{\lambda}_{\mathrm{opt}}(t)$ that
maximizes $\mathcal{F}$ for a fixed amount of time $t_{f}$. It is
clear that the formulated problem is highly non-local and, in
particular, allows for protocols which strongly deviate from the
instantaneous ground-state for intermediate times, but give a very
high final fidelity~\cite{bangbang, Chamon2011}. Here, we focus on a
more modest goal of finding protocols optimizing the instantaneous
fidelity along the path. An obvious advantage of such protocols is 
their robustness against any small changes in the couplings or shape
of pulses, especially in complex many-particle systems.


\textit{Quantum geometry.}$-$A natural way to quantify the distance
between two infinitesimally separated ground-states in Hilbert space,
is given by
$
ds^{2}
\equiv
1
- 
\abs{\bracket{\psi_{0}(\vec{\lambda})}{\psi_{0}(\vec{\lambda}+d\vec{\lambda})}}^{2}
=
g_{\mu\nu} d\lambda^{\mu}d\lambda^{\nu}
$,
where the quantum metric tensor $g_{\mu\nu}$ reads
\be
\label{eq:qgtdef}
g_{\mu\nu}
=
\mathrm{Re}\!\lb
 \bra{\psi_{0}} \overleftarrow{\pd_{\mu}} \pd_{\nu} \ket{\psi_{0}}
 -
 \bra{\psi_{0}} \overleftarrow{\pd_{\mu}} \ket{\psi_{0}}
 \bra{\psi_{0}} \pd_{\nu} \ket{\psi_{0}}
\rb,
\ee
with $\pd_{\mu}\equiv\pd/\pd\lambda^{\mu}$ and
$\bra{\psi_{0}} \overleftarrow{\pd_{\mu}} \equiv \pd_{\mu}
\bra{\psi_{0}}$.
The expansion of $ds^{2}$ in $\{d\lambda^{\mu}\}$ clearly shows that
$g_{\mu\nu}$ induces a metric on $\mathcal{M}$.
This metric tensor was first studied in~\cite{Provost1980}, and became
an object of great interest in quantum information
theory~\cite{Gu2010}, the study of quantum phase
transitions~\cite{Wang2015} and the characterization of topological
phases~\cite{topopm}.

The fact that $\mathcal{M}$ is a metric space provides us the notion
of geodesic curves.
On a Riemannian manifold, a geodesic is a path that minimizes the
distance functional
\be
\mathcal{L}(\vec{\lambda})
=
\int_{\vec{\lambda}_{i}}^{\vec{\lambda}_{f}} \!\! ds
=
\int_{0}^{t_{f}}\sqrt{g_{\mu\nu}\dot{\lambda}^{\mu}\dot{\lambda}^{\nu}}\,dt,
\ee
between two given points
$\vec{\lambda}_{i}=\vec{\lambda}(0)$ and
$\vec{\lambda}_{f}=\vec{\lambda}(t_{f})$, with
$\dot{\lambda}^{\mu} \equiv d\lambda^{\mu}/dt$.
The integrand of $\mathcal{L}$, which is 
extremized along the path, corresponds to the fidelity
$\mathcal{F}$ between infinitesimally separated
ground-states. Therefore a geodesic has the meaning of a path
maximizing the local fidelity. In Ref.~\cite{Kolodrubetz2013}, it was
shown that in the leading order of non-adiabatic response,
$(g_{\mu\nu}\dot{\lambda}^{\mu}\dot{\lambda}^{\nu})^{1/2}$ gives
the mean energy variance $\delta E$ at any particular point of the
protocol. Thus, a geodesic curve also minimizes the
energy fluctuations averaged along the path. It is interesting to
point out that the energy variance can be interpreted as the
time-component of the metric tensor,
$\delta E^2
 =
 g_{tt}
 \equiv
 \bra{\psi(t)} \overleftarrow{\pd_{t}} \pd_{t} \ket{\psi(t)} -\bra{\psi(t)} \overleftarrow{\pd_{t}} \ket{\psi(t)}
 \bra{\psi(t)} \pd_{t} \ket{\psi(t)}
$.
The equivalence between $g_{tt}$ and the energy variance follows from
$i\partial_t \ket{\psi(t)}=H \ket{\psi(t)}$. Near the adiabatic
limit, where
$\ket{\psi(t)}=\ket{\psi_0}+\mathcal{O}(\dot{\vec\lambda})$, a
geodesic can therefore also be thought of as the curve minimizing the proper
time interval along the path. While we focus on the ground-state
manifold in this Letter, these ideas apply to
excited-states as well.
Moreover, as the metric tensor has a well defined
classical limit~\cite{Kolodrubetz2016}, our findings remain valid for
classical Hamiltonian systems, where dissipation is very low.

The distance $\mathcal{L}$ along a curve is
obviously independent of the parametrization, therefore we may choose
$g_{\mu\nu} \dot{\lambda}^\mu\dot{\lambda}^\nu$ to be constant in time.
The differential equations for geodesics take then the well known
form~\cite{seesup}
\be
\label{eq:geod}
\ddot{\lambda}^{\mu}
+
\Gamma_{\nu\rho}^{\mu} \dot{\lambda}^{\nu} \dot{\lambda}^{\rho}
=
0,
\ee
where the Christoffel symbols are given by
$
\Gamma_{\nu\rho}^{\mu}
=
\frac{1}{2}
g^{\mu\xi}
\lp
\pd_{\rho} g_{\xi\nu}
+
\pd_{\nu} g_{\xi\rho}
-
\pd_{\xi} g_{\nu\rho}
\rp
$
and 
$(g^{\mu\nu})=(g_{\mu\nu})^{-1}$ is the inverse of the metric
tensor~\cite{Peterson}.
Let us highlight that the conservation of the product
$g_{\mu\nu}\dot \lambda^\mu\dot\lambda^\nu$ along a geodesic, implies
that near points where the metric tensor is large,
e.g., points corresponding to a small energy gap, the speed
$|\dot{\vec\lambda}|$ has to be small.
%
%
%
%
We note that in Ref.~\cite{Kumar2012}, geodesics were used to analyze quantum criticalities. Moreover, it has been shown that they correspond to paths minimizing the error in adiabatic and holonomic quantum computation~\cite{Rezakhani2010}. Below we illustrate how our ideas apply to two specific examples.
\begin{figure*}
\includegraphics[scale=0.54,trim=0.05mm 0.05mm 0.05mm 0.05mm, clip]{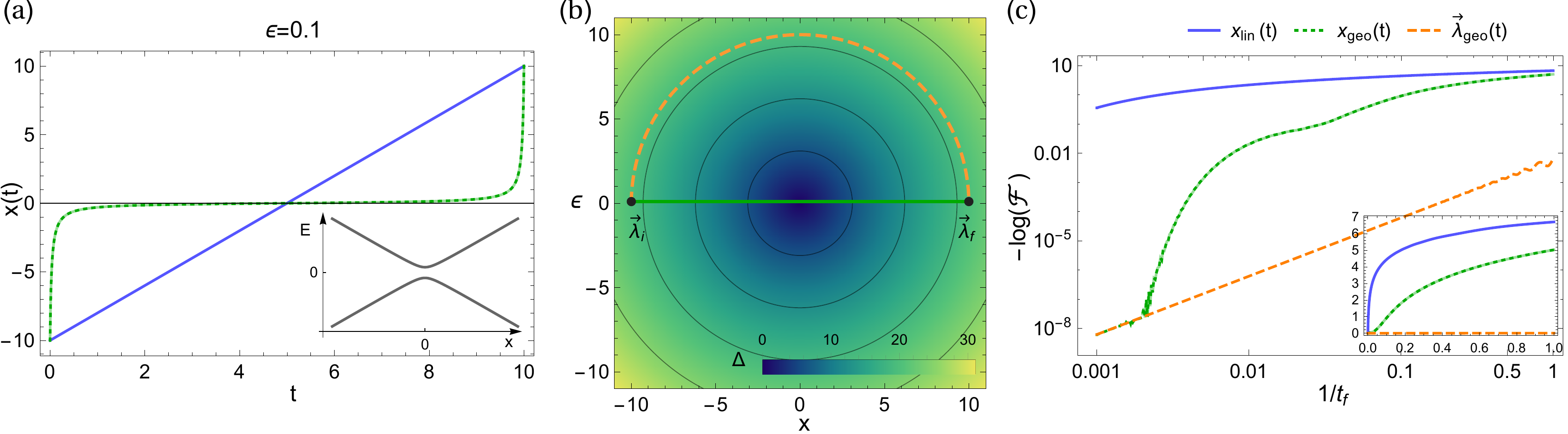}
\caption{
  (Color online)
  Geodesic for the passage through an avoided level crossing:
  (a) The linear (blue solid) and the geodesic (green dotted)
  Landau-Zener protocols are depicted for the initial $x_{i}=-10$ and
  final $x_{f}=10$ points, with $\epsilon=0.1$ and a total evolution
  time of $t_{f}=10$.
  (b) The energy gap, $\Delta = E_{1} - E_{0} = 2 \sqrt{x^{2}+\epsilon^{2}}$, is plotted in the $(x,\epsilon)$ parameter space.
  The dashed orange and the straight green lines correspond to the
  circular and constant $\epsilon$ geodesic protocols, respectively
  (see text for details).
  (c) $-\log(\mathcal{F})$ as a function of
  $1/t_{f}$, for the three
  different protocols $x_{\mathrm{lin}}(t)$ (blue solid),
  $x_{\mathrm{geo}}(t)$ 
  (green dotted) and $\vec{\lambda}_{\mathrm{geo}}(t)=\Omega_{i}(\sin\theta_{\mathrm{geo}}(t),\cos\theta_{\mathrm{geo}}(t))^{T}$ (orange
  dashed), is shown on a logarithmic scale, for the values used in
  (a). The inset shows the same on a linear scale.
  }
\label{fig:lz}
\end{figure*}


\textit{The Landau-Zener model.}$-$Let us first illustrate our formalism on
a simple two-level system given by the Landau-Zener
Hamiltonian~\cite{LZMS},
\be
H_{\mathrm{LZ}}(t)
=
x(t) \sigma^{x} + \epsilon(t) \sigma^{z}
=
\begin{pmatrix}
\epsilon(t) & x(t) \\
x(t) & -\epsilon(t)
\end{pmatrix}.
\ee
The operators $\sigma^{x}$ and $\sigma^{z}$ are the usual Pauli
matrices and
$\ket{\uparrow}=(1,0)^{T}$, $\ket{\downarrow}=(0,1)^{T}$ denote the
eigenstates of $\sigma^{z}$.
The parameter $x$ characterizes the coupling between the two levels
and $\epsilon$ the detuning.
The instantaneous eigenstates of this system are given by
\be
\label{eq:psigs}
\ket{\psi_{0,1}}
=
\mp \frac{1}{\sqrt{2}}
\frac{\Omega \mp \epsilon}{\sqrt{\Omega(\Omega \mp \epsilon)}}
\ket{\uparrow}
+
\frac{1}{\sqrt{2}}
\frac{x}{\sqrt{\Omega(\Omega \mp \epsilon)}}
\ket{\downarrow},
\ee
%
where we defined $\Omega\equiv\sqrt{x^{2}+\epsilon^{2}}$, and the
corresponding eigenenergies are $E_{0,1} = \mp \Omega$.
Our goal is to obtain the optimal control protocol
$\vec{\lambda}_{\mathrm{opt}}(t)=(x_{\mathrm{opt}}(t),\epsilon_{\mathrm{opt}}(t))^{T}$
maximizing the overlap
$\mathcal{F}(t_{f})=\abs{\bracket{\psi(t_{f})}{\psi_{0}(t_{f})}}^{2}$,
when evolving an initial ground-state $\ket{\psi_{0}(0)}$
corresponding to $\vec\lambda_i=(x_i, \epsilon)^{T}$, to the target
ground-state $\ket{\psi_{0}(t_{f})}$ corresponding to
$\vec\lambda_f=(x_f, \epsilon)^{T}$.
We assume that $\abs{x_{i,f}} \gg \epsilon$.

First, consider the simplest standard protocol, where $\epsilon$ is time-independent and $x(t)$ linearly depends on time~\cite{smoothing_note}:
$x_{\mathrm{lin}}(t) = x_{i} + (x_{f}-x_{i})  t/t_f$.
This protocol corresponds to the paradigmatic Landau-Zener
problem~\cite{Vitanov1996}, and the initial adiabatic ground-state
tunnels to the excited-state during the evolution with a finite
probability, which yields a final fidelity given by
$\mathcal{F}(t_{f})
\approx
1 - \exp
     \!\!\lb
       -\pi \frac{\epsilon^{2} t_f}{(x_{f}-x_{i})}
     \rb$.
An intuitive way to improve this protocol would be to simply adjust the speed $\dot{x}(t)$ during the evolution, slowing down near the avoided level-crossing, thereby reducing transitions to the excited-state.

Next, let us fix $\epsilon$ and consider $x(t)$ as an arbitrary time-dependent parameter in the system. The quantum metric tensor is very easy to compute using the ground-state~\eqref{eq:psigs}:
\be
g_{xx}=\frac{\epsilon^{2}}{4(x^{2} + \epsilon^{2})^{2}},
\ee
and thus the geodesic protocol, determined by $g_{xx}\dot x^2=$const, reads
$
x_{\mathrm{geo}}(t) = \epsilon \tan \lb \alpha_{i} + (\alpha_{f}-\alpha_{i}) \, t/t_f \rb,
$
where
$\alpha_{i,f} = \arctan(x_{i,f}/\epsilon)$. 
The geodesic protocol slows down close to the avoided
level-crossing (Fig.~\ref{fig:lz}(a)), and hence minimizes the tunneling probability to the
excited-state during the evolution (Fig.~\ref{fig:lz}(c)). In the context of quantum adiabatic search algorithms, a similar protocol is discussed in Ref.~\cite{Cerf2002}, obtained by enforcing
adiabatic evolution on each infinitesimal time interval.
In Ref.~\cite{Bason2012}, such protocol was implemented experimentally,
using a two-level quantum system consisting of Bose-Einstein condensates
in optical lattices, achieving a higher fidelity than a linear driving
protocol.

It is intuitively clear that one can further optimize the protocol by
increasing the number of control parameters. Mathematically, this is
reflected in the fact that by choosing the parameter manifold, we
consider a subset of the full Hilbert space. Thus, the geodesics found
within this manifold will generally have non-vanishing geodesic
curvature. By introducing extra parameters, i.e., by increasing the
dimensionality of the subset, we can find geodesics with smaller and
smaller geodesic curvature, which correspond to
shorter geodesics and
hence higher final fidelity. In the example discussed here, the
geodesic we found has zero geodesic curvature, so introducing more
parameters will not affect the length. To illustrate this, let us
expand the parameter manifold and allow both $x$ and $\epsilon$ to
depend on time: $\vec{\lambda}(t) = (x(t),\epsilon(t))^{T} = \Omega(t)
\, (\sin\theta(t),\cos\theta(t))^{T}$. In coordinates $\mu,\nu \in \{
\Omega, \theta \}$, the quantum metric tensor shortens to
$(g_{\mu\nu})=\diag(0,1/4)$. Obviously, the metric tensor has zero
components with respect to $\Omega$, as changing the overall energy scale does not affect the eigenstates. In turn, this implies that we are free to choose the arbitrary protocol $\Omega(t)$.
Let us choose the circular protocol $\Omega_{\mathrm{geo}}(t)=\Omega_{i}$.
The geodesic equation for $\theta(t)$ reduces then to
$\ddot{\theta}=0$, which yields
$\theta_{\mathrm{geo}}(t)=\theta_{i} + (\theta_{f}-\theta_{i}) \,
t / t_{f}$,
with $\theta_{i,f}=\arctan(x_{i,f}/\epsilon_{i,f})$.
This protocol $\vec{\lambda}_{\mathrm{geo}}(t)$ is
nothing but a great circle in the full $SU(2)$ manifold of the
two-level system, and thus has zero geodesic curvature. Therefore
introducing the only remaining independent parameter $\phi$, which
defines the magnetic field angle in the $xy$-plane, will not affect
the geodesic~\cite{seesup}. It is easy to see that the circular
protocol is equivalent to the one with constant $\epsilon$, discussed earlier, up to an overall rescaling of $\Omega$.  

Despite the formal equivalence between the constant $\epsilon$ and
circular geodesic protocols leading to the same distance, there is an
important physical difference between them. In the limit of small
$\epsilon$, the former protocol corresponds to crossing a small gap
region, while the latter corresponds to the time-independent gap
(Fig.~\ref{fig:lz}(b)). The slightly counterintuitive equivalence between
these two geodesic protocols is hidden in their very different velocity
profiles. In the former case, one first moves very fast to the small
gap region $x\sim \epsilon$ and then slowly crosses it. In the latter
case, one changes $\theta$ with a uniform velocity without changing the
gap. It is intuitively clear that the circular protocol is more robust
against introducing additional degrees of freedom, e.g., introducing a
third excited-state. These extra degrees of freedom should also break
the degeneracy between the geodesics. 
%
Even in the two-level case the circular protocol generally performs
better, since the adiabatic approximation breaks down at much smaller
velocities for the constant $\epsilon$-protocol.
Except for very large $t_f$, where the two protocols are equivalent,
they give the same fidelity (c.f.~green and orange lines in
Fig.~\ref{fig:lz}(c)).
\begin{figure*}[]
\includegraphics[scale=0.62,trim=0.01mm 4.1mm 0.01mm 0.01mm, clip]{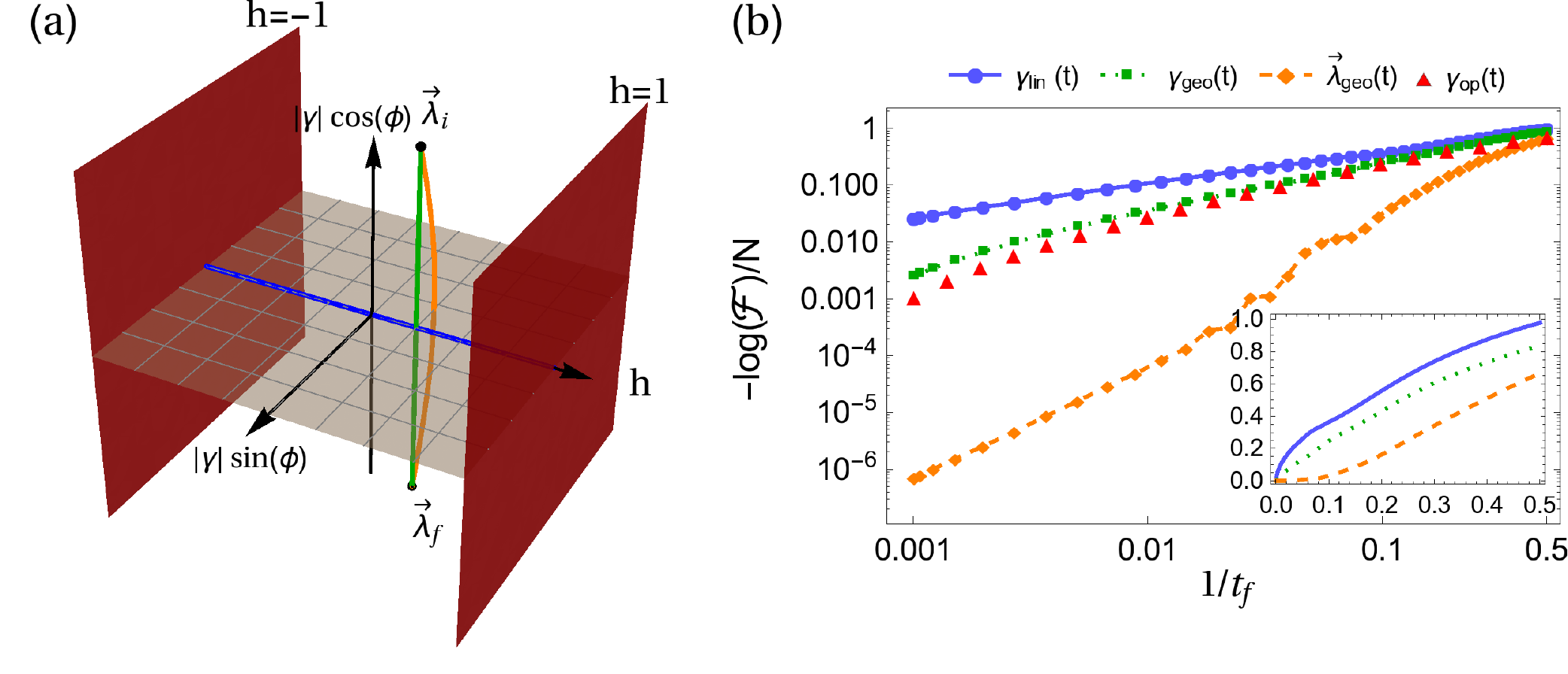}
\caption{(Color online) Geodesic passage through a quantum phase transition:
(a) The phase diagram of the rotated XY spin chain in cylindrical
coordinates $(\abs{\gamma} \cos\phi, \abs{\gamma} \sin\phi, h)$ is depicted.
The two red planes ($\abs{h}=1$) indicate the Ising criticality, 
where the system undergoes a continuous transition between a paramagnetic
and a ferromagnetic phase.
The blue line ($\gamma=0$) marks the anisotropic transition,
separating the two different aligned ferromagnetic phases. 
The green and orange lines illustrate the driving protocols for
crossing and avoiding the quantum criticality, respectively.
(b) $-\log(\mathcal{F})/N$ for the three
different driving protocols is shown on a logarithmic scale, with $\gamma_{i}=1$,
$\gamma_{f}=-1$, $h=0.5$ and $N=900$. The inset shows the same on a linear scale.
For comparison we also plotted $-\log(\mathcal{F})/N$ for the optimal power-law
protocol
$\gamma_{\mathrm{op}}(t)=\sgn[\gamma_{\mathrm{lin}}(t)]\abs{\gamma_{\mathrm{lin}}(t)}^{r_{\mathrm{op}}}$
(red triangles).
}
\label{fig:xyg}
\end{figure*}



\textit{The anisotropic XY spin chain.}$-$Let us now apply our
analysis to a quantum many-body system.
For this purpose, we consider the illustrative example of the
anisotropic XY spin chain in a transverse magnetic field~\cite{XYlit},
given by the Hamiltonian 
\be
\label{eq:xyham}
H_{XY}
=
-
\sum_{j=1}^{N}
\lb
\frac{1+\gamma}{2} \, \sigma_{j}^{x} \sigma_{j+1}^{x}
+
\frac{1-\gamma}{2} \, \sigma_{j}^{y} \sigma_{j+1}^{y}
+
h \, \sigma_{j}^{z}
\rb,
\ee
where $\sigma_{j}^{\alpha}$, with $\alpha=x,y,z$, are the Pauli
matrices describing the spin on the $j-$th site of the chain.
We assume periodic boundary conditions,
$\sigma_{N+1}^{\alpha}=\sigma_{1}^{\alpha}$, and fix the overall
energy scale to unity.
The parameters of the model are the anisotropy $\gamma$ of the nearest
neighbor spin-spin exchange interaction along the $x$ and $y$
direction, and the transverse magnetic field $h$.
We add an additional tuning parameter $\phi$, describing a
simultaneous rotation of all spins around the $z$ axis by an angle
$\phi/2$.
The corresponding Hamiltonian is obtained by 
$\tilde{H}_{XY}(h,\gamma,\phi) = R_{z}(\phi) H_{XY} R_{z}^{\dag}(\phi)$, 
where $R_{z}(\phi) = \prod_{l=1}^{N}\exp(-i\frac{\phi}{2}\sigma_{l}^{z})$.
We note that such a rotation of the whole system by $\phi$ does not affect its 
spectrum, but it modifies the eigenstates.
%
%
%
$\tilde{H}_{XY}$ can be mapped to non-interacting fermions
using the standard Jordan-Wigner and Fourier
transformations~\cite{Dutta2015}, providing a unique ground-state
$\ket{\mathrm{GS}}$ throughout the entire parameter space, once a
particular fermion parity is fixed~\cite{Lieb1961}.
The phase diagram of the model is illustrated in Fig.~\ref{fig:xyg}(a). 

We focus on the $\abs{h} \leq 1$ region of the parameter space
and study the fidelity
$\mathcal{F}(t_{f}) = \abs{\bracket{\psi(t_{f})}{\mathrm{GS}(t_{f})}}^{2}$ 
for the preparation of a target ground-state
$\ket{\mathrm{GS}(t_{f})}$ from an initial ground-state
$\ket{\mathrm{GS}(0)}$, lying in a different phase region than the
target state.
Let us analyze the passage through the anisotropic
transition line with fixed $h=0.5$, for the initial $\gamma_{i}=1$ and
final $\gamma_{f}=-1$ points.
As in the previously studied case, we compare the three different
protocols: a linear $\gamma_{\mathrm{lin}}(t)$, a geodesic
$\gamma_{\mathrm{geo}}(t)$ and a geodesic one which avoids the quantum
phase transition 
$\vec{\lambda}_{\mathrm{geo}}(t)=\abs{\gamma(t)}(\cos\phi(t), \sin\phi(t))^{T}$.
The corresponding quantum metric tensor was calculated in
Ref.~\cite{Zanardi2007, Kolodrubetz2013} and reads
\be
g_{\gamma\gamma} = \frac{1}{16\abs{\gamma}(1+\abs{\gamma})^{2}},
\quad
g_{\phi\phi} = \frac{\abs{\gamma}}{8(\abs{\gamma}+1)},
\quad
g_{\gamma\phi} = 0.
\ee

In Fig.~\ref{fig:xyg}(b), we plot the resulting final fidelities. 
For the linear protocol
$\gamma_{\mathrm{lin}}(t)=\gamma_{i}+(\gamma_{f}-\gamma_{i})t/t_f$,
the observed result, $-\frac{1}{N}\log(\mathcal{F}) \sim 1/\sqrt{t_f}$,
is in perfect agreement with the general Kibble-Zurek predictions for linear
quenches~\cite{Polkovnikov2005}.
Crossing the quantum phase transition along a geodesic protocol
clearly yields a higher final fidelity, as can be seen in
Fig.~\ref{fig:xyg}(b).
The energy gap vanishes at the quantum phase transition $\gamma=0$ and
consequently the metric diverges, which imposes
$\dot{\gamma}_{\mathrm{geo}} \to 0$ on the velocity when approaching
the transition.
This is due to the fact that the product
$g_{\gamma\gamma}\,\dot{\gamma}^{2}$
has to be constant along a geodesic.
%
The corresponding geodesic takes then the form
$
\gamma_{\mathrm{geo}}(t)
=
\sgn[X(t)] \tan^{2}[X(t)]
$,
where 
$X(t)=\chi_{i} + (\chi_{i}-\chi_{f}) t/t_{f}$
and
$\chi_{i,f} = \sgn(\gamma_{i,f})\arctan(\sqrt{\abs{\gamma_{i,f}}})$.
%
%
In Ref.~\cite{Barankov2008, Chiara2013}, the optimal adiabatic
crossing of a quantum critical point has been analyzed.
More specifically, they found that in order to minimize the number of
excitations, the driving protocol should be given by a power-law,
where the exponent serves as a minimization parameter.
However, this optimization of the exponent yields only an incremental
improvement of the final fidelity compared to the geodesic protocol
(see~Fig.~\ref{fig:xyg}(b)). And thus the geodesic still gives a
nearly optimal protocol despite the breakdown of the adiabatic
perturbation theory.

%
%
%
%
%
%
%
%

Finally, let us study the final fidelity when tuning both $\gamma$ and
$\phi$ simultaneously.
In this case, the metric tensor can be expressed by
$g_{\mu\nu} = \frac{1}{4} \diag(1, \sin^{2}\eta)$,
where $\mu,\nu \in \{ \eta, \varphi \}$,
defined by $\gamma=\tan^{2}\eta$ and $\phi=\sqrt{2}\,\varphi$.
The resulting geodesic protocol $\vec{\lambda}_{\mathrm{geo}}(t)$
is thus given by a great circle on the sphere defined by $\{ \eta, \varphi\}$.
This geodesic protocol gives significantly better final fidelity than
the linear one as it avoids the critical point (c.f.~Fig.~\ref{fig:xyg}(b)).


\textit{Conclusion}.$-$We used a geometric approach to
obtain optimal protocols for the adiabatic preparation of
ground-states in quantum many-body systems close to the adiabatic limit.
Those are geodesics in the space of control parameters, maximizing the
overlap between the evolved state and the target state, while
simultaneously keeping the quantity
$g_{\mu\nu}\dot{\lambda}^{\mu}\dot{\lambda}^{\nu}$, which is equal to the
energy variance, stationary along the path.
Further, we showed that by increasing the number of control parameters
and tuning them along geodesic paths on the extended parameter space
can provide a further increase in the final fidelity. This method can be applied to various optimization problems like finding best quantum annealing protocols, optimum adiabatic path for quantum simulation or minimization of heating in experiments with ultra-cold atoms.

\begin{acknowledgments}
\label{Section: Acknowledgments}
\textit{Acknowledgments.}$-$
The authors thank A.~Dunsworth, V.~Gritsev, M.~Kolodrubetz, and
P.~Roushan for enlightening discussions. This work was supported by 
AFOSR FA9550-13-1-0039, NSF
DMR-1506340 (T.~S. and A.~ P.), ARO W911NF1410540 (M.~T. and A.~P.) and the Swiss National Science Foundation (SNSF).
\end{acknowledgments}

\newpage
\clearpage
\onecolumngrid

\renewcommand{\theequation}{S\arabic{equation}} 
\renewcommand{\thepage}{S\arabic{page}} 
\renewcommand{\thesection}{S\arabic{section}}  
\renewcommand{\thetable}{S\arabic{table}}  
\renewcommand{\thefigure}{S\arabic{figure}}
\renewcommand{\bibnumfmt}[1]{[{\normalfont S#1}]}
\setcounter{page}{0}
\setcounter{equation}{0}


\begin{center}
{\bf \Large Supplementary Material for the paper \\ ``Geodesic paths for quantum many-body systems''} \\ ~ \\ by M. Tomka, T. Souza, S. Rosenberg, and A. Polkovnikov
\end{center}

\section{Quantum Geometric Tensor}
\label{sec:sup1}

In this section we show that the quantum metric tensor $g_{\mu\nu}$,
introduced in the main text, is the symmetric part of the more general
{\it quantum geometric tensor}.
The quantum geometric tensor was introduced by Provost and
Vallee~\cite{ProvostVallee1980}, but the term itself first 
appeared in a work from M.~Berry~\cite{BerryinSW1989}.
For a ground-state $\ket{\psi_{0}}$ of a generic quantum system, it
is given by 
\be
\label{eq:QGT}
\chi_{\mu\nu}
\equiv
\bra{\psi_{0}} \overleftarrow{\pd_{\mu}} \pd_{\nu} \ket{\psi_{0}}
-
\bra{\psi_{0}} \overleftarrow{\pd_{\mu}} \ket{\psi_{0}}
\bra{\psi_{0}} \pd_{\nu} \ket{\psi_{0}}.
\ee
Alternatively, it can also be expressed as a sum over all the
eigenstates $\ket{\psi_{m}}$, by
\be
\chi_{\mu\nu}
=
\sum_{m \neq 0}
\frac{\bra{\psi_{0}}\pd_{\mu}\op{H}\ket{\psi_{m}}\bra{\psi_{m}}\pd_{\nu}\op{H}\ket{\psi_{0}}}{(E_{0}-E_{m})^{2}},
\ee
where the resolution of identity
$\sum_{m}\ket{\psi_{m}}\bra{\psi_{m}}=\op{1}$ and $\bra{\psi_{m}}\pd_{\mu}\ket{\psi_{n}}=\bra{\psi_{m}}\pd_{\mu}\op{H}\ket{\psi_{n}}/(E_{n}-E_{m})$,
valid for $m \neq n$, were used.

The symmetric part of the quantum geometric tensor
\be
g_{\mu\nu}
\equiv
\frac{1}{2} (\chi_{\mu\nu} + \chi_{\nu\mu})
=
\mathrm{Re}(\chi_{\mu\nu}),
\ee
corresponds to the quantum metric tensor used in the main text.
It defines a Riemannian metric in the parameter manifold
$\mathcal{M}$ with respect to the local coordinates
$\{\lambda^{\mu}\}$, and therefore also a measure of distances between
different ground-states, identified as points in $\mathcal{M}$ by the
map $\lp \lambda^{\mu} \rp \in \mathcal{M} \longleftrightarrow
\ket{\psi_{0} (\lambda^{\mu}) }$.
The distance $d s$ between two ground-states differing by an
infinitesimal variation of coordinates in $\mathcal{M}$ is then given
by
\be
\label{eq:ds-in-manifold}
d s^2
\equiv
1 -
\abs{\bracket{\psi_{0}(\vec{\lambda})}{\psi_{0}(\vec{\lambda}+d\vec{\lambda})}}^{2}
=
g_{\mu \nu} d \lambda^{\mu} d \lambda^{\nu},
\ee
where Einstein summation convention over repeated indices is implied.

The anti-symmetric part of the quantum geometric tensor defines the
Berry curvature
\be
F_{\mu\nu}
\equiv
i (\chi_{\mu\nu} - \chi_{\nu\mu})
=
-2\, \mathrm{Im}(\chi_{\mu\nu}),
\ee
which gives rise to the Berry phase and a topological invariant known
as the first Chern number.

\section{Relationship between the quantum metric tensor and the energy variance}
\label{sec:sup2}

In the following, we present the relationship between the energy
fluctuations $\delta E$ and the quantum metric tensor $g_{\mu\nu}$.
The energy fluctuations are defined by
\be
\delta E^{2}(t)
\equiv
\bra{\psi(t)}H^{2}\ket{\psi(t)} - \bra{\psi(t)}H\ket{\psi(t)}^{2}.
\ee
Within adiabatic perturbation theory~\cite{DeGrandi-Polkovnikov-2010},
we can compute $\ket{\psi(t)}$ in powers of the driving velocities
$\dot{\lambda}^{\mu}$
\be
\ket{\psi(t_{f})}
=
\ket{\psi_{0}}
-
i \dot{\lambda}^{\mu} \sum_{m \neq 0} \frac{\bra{\psi_{m}} \pd_{\mu} H \ket{\psi_{0}}}{(E_{m}-E_{0})^{2}} \ket{\psi_{m}}
+
\dots,
\ee
where we assumed that $\abs{\dot{\vec{\lambda}}} \ll 1$.
Inserting this expansion into the expression of the energy
fluctuations yields
\be
\delta E^{2}
\approx
\chi_{\mu\nu} \dot{\lambda}^{\mu} \dot{\lambda}^{\nu}
=
\lp g_{\mu\nu} - \frac{i}{2}\, F_{\mu\nu} \rp \dot{\lambda}^{\mu} \dot{\lambda}^{\nu}
=
g_{\mu\nu} \dot{\lambda}^{\mu} \dot{\lambda}^{\nu},
\ee
showing that the metric tensor defines the leading non-adiabatic
correction of the energy fluctuations
$
\delta E^{2}
\approx
g_{\mu\nu} \dot{\lambda}^{\mu} \dot{\lambda}^{\nu}
$.
This result was shown in Ref.~\cite{Kolodrubetz2013sup}.
We note that due to energy conservation in a closed quantum
system, the energy fluctuations are equal to the fluctuations
of the work done on the system, $\delta W^{2}$, and therefore the
quantum metric tensor can also be measured through the work
fluctuations.

%

\newpage
\section{Geodesics and the Euler-Lagrange Equations}

In general, the quantum distance between two ground-states situated at
$\vec{\lambda}_{i}$ and $\vec{\lambda}_{f}$, connected by a
path $\vec{\lambda}$ in parameter space, can be written as
\be
\mathcal{L}(\vec{\lambda}) = \int_{\vec{\lambda}_{i}}^{\vec{\lambda}_{f}} d s 
=
\int_{\vec{\lambda}_{i}}^{\vec{\lambda}_{f}} \sqrt{g_{\mu\nu} \, d\lambda^{\mu} \, d\lambda^{\nu}}.
\ee
The curve $\vec{\lambda}$ may be parametrized by $t$ such that 
$\vec{\lambda} \equiv \vec{\lambda} (t)$ with
$\vec{\lambda} (t_{i}) = \vec{\lambda}_{i}$ and
$\vec{\lambda} (t_{f}) = \vec{\lambda}_{f}$,
and consequently the previous equation becomes
\be
\label{eq:dist-functional-1}
\mathcal{L} (\vec{\lambda}) 
= 
\int_{t_{i}}^{t_{f}}
\sqrt{
 g_{\mu\nu} \, \frac{d\lambda^{\mu}}{d t} \, \frac{d\lambda^{\nu}}{d t}
} \, d t 
\, . 
\ee
%
We note that the above functional $\mathcal{L}$ is invariant under any
reparametrization of the parameter $t$, i.e.,
$\tau = \tau(t)$.
Further, the stationary curve $\vec{\lambda}_{\mathrm{geo}}(t)$ of $\mathcal{L}$
will naturally inherit this property and is referred to as the
{\it geodesic} connecting the boundary points.
In case $t_{i}=0$, a convenient parametrization is
$t = t_{f} \, \tau $, $d t = t_{f} \, d\tau$,
and then Eq.~(\ref{eq:dist-functional-1}) becomes
\be
\label{eq:length-action}
\mathcal{L}(\vec{\lambda})
=
\int_{0}^{1} \sqrt{g_{\mu\nu} \, \frac{d\lambda^{\mu}}{d \tau} \,
  \frac{d\lambda^{\nu}}{d \tau}} \, d \tau \, .
\ee
Such choice is often referred to as the {\it proper parametrization}.



The path $\vec{\lambda}_{\mathrm{geo}}(t)$ that minimizes the distance
between the fixed endpoints is obtained by 
\be
\frac{\delta \mathcal{L}}{\delta \vec{\lambda}} = 0.
\ee
%
%
%
%
%
It turns out, the variation of $\mathcal{L}(\vec{\lambda})$ can be
calculated in an easier way. To this end, let us introduce the 
action functional $\mathcal{E}$, defined by
\be
\label{eq:energy-action}
\mathcal{E} = \frac{1}{2} \int_{0}^{1} \lp g_{\mu\nu} \, \frac{d\lambda^{\mu}}{d \tau} \, \frac{d\lambda^{\nu}}{d \tau} \rp \, d \tau,
\ee
which has a much simpler integrand.
The Cauchy-Schwarz inequality for square-integrable functions,
\be
\lp \int_{a}^{b} f(t) \, h(t) \, d t \rp^{2} \leq \lp \int_{a}^{b} f^{2}(t) \, d t  \rp \lp \int_{a}^{b} h^{2}(t) \, d t  \rp,
\ee
for $f(t) = 1$, $h(t) = \sqrt{ g_{\mu \nu} \lp d\lambda^{\mu}/d\tau \rp \lp d\lambda^{\nu}/d\tau \rp }$, $a=0$ and $b=1$, implies then that
\be
\big(\mathcal{L}\,\big)^{2} \leq 2\, \mathcal{E},
\ee
where the equality holds if and only if $h$ is
constant.
Hence, if we apply the principle of stationary action to the
functional $\mathcal{E}$, we also obtain the stationary
solutions of $\mathcal{L}$, with one very important caveat: the
functional $\mathcal{E}$ is not invariant under change of
parametrizations, as one can easily verify by
Eq.~(\ref{eq:energy-action}).
Consequently, the stationary curve of $\mathcal{E}$ will also be
stationary for $\mathcal{L}$, provided that the solution curve
$\vec{\lambda}_{\mathrm{geo}}(t)$ is parametrized only by linear functions of
$t$~\cite{Spivak-Collection-Vol-1, Morse-Theory-1963}.

The stationary solutions of $\mathcal{E}$ are then found by the
standard procedure~\cite{Spivak-Collection-Vol-1}, and follow from the 
Euler-Lagrange equations, which in local coordinates
$\{\lambda^{\mu}\}$ read
\be
\label{eq:geod-eq}
\frac{d^{2}\lambda^{\mu}}{d \tau^{2}}
+
\Gamma^{\mu}_{\nu\rho}
\frac{d\lambda^{\nu}}{d\tau}
\frac{d\lambda^{\rho}}{d\tau}
=
0,
\ee
where $\Gamma_{\mu}^{\nu\rho}$ are the Christoffel symbols of the
second kind, defined by
\be
\Gamma_{\nu\rho}^{\mu}
=
\frac{1}{2}
g^{\mu\xi}
\lp
\pd_{\rho} g_{\xi\nu}
+
\pd_{\nu} g_{\xi\rho}
-
\pd_{\xi} g_{\nu\rho}
\rp,
\ee
with $\pd_{\mu} \equiv \pd/\pd \lambda^{\mu}$
and
$g^{\mu\xi}$ are the components of the inverse of the metric tensor
$g_{\mu\xi}$, i.e.,
$(g^{\mu\xi}) = (g_{\mu\xi})^{-1}$.

We note that the integrand of $\mathcal{L}$ corresponds to
$1-\mathcal{F}$, for infinitesimally separated ground-states, as can
be seen from Eq.~(\ref{eq:ds-in-manifold}). 
Therefore, geodesics are paths maximizing the local fidelity.
Moreover, the integrand of $\mathcal{E}$ gives the energy
fluctuations $\delta E$ in the leading order of non-adiabatic
response~\cite{Kolodrubetz2013sup},
at any particular point of the protocol. Thus, geodesics are curves
that also minimizes energy fluctuations averaged along the path.

\newpage
\section{Additional parameter space extension for the two-level system}
\label{sec:geoxandy}

In this section we compute the geodesics when tuning the magnetic field in
the xy-plane of the two-level system.
Let us consider the Hamiltonian
\be
H
=
\begin{pmatrix}
\epsilon & x(t) -i y(t) \\
x(t) + i y(t) & -\epsilon
\end{pmatrix}.
\ee
We will use the following coordinates to simplify the calculations
\be
x(t) = h(t) \cos \phi(t),
\qquad
y(t) = h(t) \sin \phi(t),
\ee
with the inverse given by
\be
h^{2}(t) = x^{2}(t) + y^{2}(t),
\qquad
\tan\phi(t) = \frac{y(t)}{x(t)}. 
\ee
The Hamiltonian reduces therefore to
\be
H=
\begin{pmatrix}
\epsilon & h(t) e^{-i\phi(t)} \\
h(t) e^{i\phi(t)} & -\epsilon
\end{pmatrix}.
\ee
The eigenstates are given by
%
\be
\ket{\psi_{0,1}}
=
\mp \frac{1}{\sqrt{2}} \sqrt{1\mp\frac{\epsilon}{\sqrt{h^{2}+\epsilon^{2}}}} \ket{\uparrow}
+
\frac{1}{\sqrt{2}} \frac{h \, e^{i\phi}}{\sqrt{(h^{2}+\epsilon^{2})\mp\epsilon\sqrt{h^{2}+\epsilon^{2}}}} \ket{\downarrow},
\ee
with the corresponding eigenenergies $E_{0,1}= \mp \sqrt{h^{2}+\epsilon^{2}}$.
The metric tensor with respect to $h$ and $\phi$ reads
\be
(g_{\mu\nu})
=
\begin{pmatrix}
g_{h h} & g_{h \phi} \\
g_{\phi h} & g_{\phi \phi}
\end{pmatrix}
=
\begin{pmatrix}
\frac{1}{4} \frac{\epsilon^{2}}{(h^{2}+\epsilon^{2})^{2}} & 0 \\
0 & \frac{1}{4} \frac{h^{2}}{(h^{2}+\epsilon^{2})}
\end{pmatrix}.
\ee

We note that for $\epsilon = 0$, the metric element $g_{hh}$ equals to
$0$ and hence we are left with a pseudo-Riemannian metric.
This is a consequence of the fact that the ground-state
$\ket{\psi_{0}}$ is independent of $h$ for
$\epsilon=0$ and therefore there is no notion of distance along the
$h$ direction.
Consequently, we are free to choose $h(t)$, such that $h>0$, since we
want to avoid the degeneracy point $E_{0}=E_{1}$. The most simple
function $h(t)$ satisfying this is the constant one.
The remaining geodesic equation for $\phi(t)$, obtained by lowering
the index in the geodesic equations (Eq.~(\ref{eq:geod}) of the main text) in
order to avoid the use of the inverse metric, is then simply
\be
\phi'' = 0,
\ee
with the solution
$\phi(t)=(\phi_{f}-\phi_{i})\frac{t}{t_{f}}+\phi_{i}$.

Let us come back to the case $\epsilon \neq 0$.
We observe that when rescaling $h=\epsilon \tilde{h}$, the metric can be
simplified to
\be
(g_{\mu\nu})
=
\begin{pmatrix}
g_{\tilde{h} \tilde{h}} & g_{\tilde{h} \phi} \\
g_{\phi \tilde{h}} & g_{\phi \phi}
\end{pmatrix}
=
\begin{pmatrix}
\frac{1}{4} \frac{1}{(\tilde{h}^{2}+1)^{2}} & 0 \\
0 & \frac{1}{4} \frac{\tilde{h}^{2}}{(\tilde{h}^{2}+1)}
\end{pmatrix}.
\ee
This metric simplifies even further, by introducing
\be
\tilde{h}(t) = \tan\vartheta(t),
\ee
and hence we obtain
\be
(g_{\mu\nu})
=
\begin{pmatrix}
g_{\vartheta \vartheta} & g_{\vartheta \phi} \\
g_{\phi \vartheta} & g_{\phi \phi}
\end{pmatrix}
=
\begin{pmatrix}
\frac{1}{4}  & 0 \\
0 & \frac{1}{4} \sin^{2}\vartheta
\end{pmatrix}.
\ee
The corresponding Christoffle symbols are
\begin{align}
&
\Gamma_{\vartheta\vartheta}^{\vartheta} = 0,
\quad
\Gamma_{\vartheta\phi}^{\vartheta} = 0,
\quad
\Gamma_{\phi\vartheta}^{\vartheta} = 0,
\quad
\Gamma_{\phi\phi}^{\vartheta} = -\cos\vartheta\,\sin\vartheta,
\nonumber \\
&
\Gamma_{\vartheta\vartheta}^{\phi} = 0,
\quad
\Gamma_{\vartheta\phi}^{\phi} = \cot\vartheta,
\quad
\Gamma_{\phi\vartheta}^{\phi} = \cot\vartheta,
\quad
\Gamma_{\phi\phi}^{\phi} = 0,
\end{align}
and the geodesic equations are given by
\be
- \cos\vartheta \, \sin\vartheta \, \phi'^{2} + \vartheta'' = 0,
\qquad
2 \cot\vartheta \, \vartheta' \phi' + \phi'' = 0.
\ee
The resulting geodesics are thus great circles on the sphere defined
by the coordinates $\{ \vartheta, \phi \}$. A detailed derivation
of this is given in~\cite{Nakahara}.

\newpage



\begin{thebibliography}{99}
\bibitem{smoothing_note} To avoid strong non-adiabatic effects related to initial and final transients, we always deal with protocols sufficiently smoothened around the initial and final times. We checked that all our results are insensitive to the details of the smoothing procedure.
\end{thebibliography}
\end{document}